\documentclass[aip,jcp,amsmath,amssymb,preprint]{revtex4-2}
\usepackage{graphicx}
\usepackage{mathrsfs}
\usepackage{bm}
\usepackage{xcolor}

\newcommand{\ket}[1]{\ensuremath{| #1 \rangle}}
\newcommand{\bra}[1]{\ensuremath{\langle #1 |}}

\newcommand{\ii}{\ensuremath{\text{i}}}
\newcommand{\ud}{\ensuremath{\text{d}}}
\newcommand{\vect}[1]{{\bf #1}}

\newcommand{\tr}{\ensuremath{\text{Tr}}}

\newcommand{\re}{\ensuremath{\text{Re}}}
\newcommand{\im}{\ensuremath{\text{Im}}}
\DeclareMathOperator{\arcsinh}{arcsinh}

\begin{document}

\title{Double-quantum spectroscopy of dense atomic vapors: interplay between Doppler and self-broadenings}

\author{Cyril Falvo}
\email{cyril.falvo@universite-paris-saclay.fr}
\affiliation{Université Paris-Saclay, CNRS, Institut des Sciences Moléculaires d'Orsay, 91405, Orsay, France}
\affiliation{Université Grenoble-Alpes, CNRS, LIPhy, 38000 Grenoble, France}
\author{Hebin Li}
\affiliation{Department of Physics, Florida International University, Miami, Florida 33199, USA}
\date{\today}

\begin{abstract}
In this article, we present a simulation study of the linear and nonlinear spectroscopy of dense atomic vapors. Motivated by recent experiments, we focus on double quantum spectroscopy which directly probes dipole-dipole interactions. By including explicitly thermal velocity, we show that temperature has an important impact on the self-broadening mechanism of the linear and nonlinear spectra. We also provide analytical expressions for the response functions in the short time limit using the two-body approximation which shows that double quantum spectroscopy for atomic vapors directly probes the transition amplitude of the electronic excitation between two atoms.  We also propose an expression for the double quantum spectrum that includes the effect of Doppler broadening and we discuss the effect of density on the spectrum. We show that when Doppler broadening is negligible compared to self-broadening the double quantum spectrum scales with the atomic density while when Doppler broadening dominates it scales as the square of the density.
\end{abstract}

\maketitle

\section{Introduction}
Resonant dipole-dipole interactions control many processes in atomic and molecular physics.  For example, they are responsible for vibrational and electronic exciton transport in biological systems,\cite{Moore:1975vn,Torii:1992uq} in molecular aggregates,\cite{Spano:2010kq} or in photosynthetic units.\cite{Thyrhaug:2018th} In atomic physics, dipole-dipole interactions play a  fundamental role, e.g. in optical atomic clocks,\cite{Swallows.:2011wv,Cidrim:2021uu} and atom-based quantum simulators.\cite{Browaeys:2016vy} Resonant dipole-dipole interactions are also important for the optical properties of atomic vapors as it contributes to shifts and broadenings of the absorption lines including the so-called self-broadening.\cite{ukaszewski:1983tp,Maki:1991aa,Sautenkov:1996wa,Guo:1996ts,Kampen:1999aa,Kondo:2006wc,Siddons:2008wj,Li:2008aa,Li:2009tz,Weller:2011vr,Hanley:2015aa,Keaveney:2012tw,Silans:2018ve,Peyrot:2019wy} In atomic vapors different processes control the broadening mechanism and the spectroscopic response. These processes strongly depend on the density and temperature of the system. At very low temperature and density, the broadening is dominated by both the natural linewidth $\Gamma_0$ and Doppler broadening $\Gamma_\text{D}$. As increasing temperature and density, the effect of self-broadening $\Gamma_{\text{self}}$ resulting from dipole-dipole interactions between atoms increases and eventually dominates for dense vapors. As a result for dense vapors, above the Doppler-broadening regime, self-broadening can be directly observed in absorption or in reflection through a linewidth linearly dependent on the density.\cite{Kondo:2006wc,Siddons:2008wj,Li:2008aa,Li:2009tz,Weller:2011vr,Hanley:2015aa}\par
Coherent multidimensional spectroscopy such as two-dimensional coherent spectroscopy (2DCS) is a powerful tool to characterize the electronic dynamics and self-broadening in atomic vapors.\cite{Lorenz:2005wh,Lorenz:2008aa,Dai:2010aa,Dai:2012qf,Gao:2016aa,Yu:2019aa,Liang:2021uj,Lomsadze:2018aa,Lomsadze:2020aa,Binz:2020vi,Bruder:2015vk,Li:2017ab,Bruder:2019aa}  For example, spectral fluctuations in dense atomic vapors have been probed using three-pulse photon echo spectroscopy.\cite{Lorenz:2005wh,Lorenz:2008aa} In addition, two-dimensional double quantum spectroscopy (DQS)  has been used to show the collective dynamics in atomic vapors that result from dipole-dipole interactions.\cite{Dai:2012qf,Gao:2016aa,Yu:2019aa,Liang:2021uj,Lomsadze:2018aa,Lomsadze:2020aa}
DQS,  which probes the two-exciton manifold, was first introduced as a direct measure of electron correlation in molecular systems.\cite{Mukamel:2007aa,Yang:2008ac,Kim:2009aa,Kim:2009ab,Nemeth:2010aa,Christensson:2010aa,Fingerhut:2012aa,Konold:2015aa,Dostal:2018aa,Mueller:2018aa} In the case of atomic vapors, the DQS signal completely vanishes for uncoupled atoms and is therefore a direct signature of dipole-dipole interactions.\cite{Dai:2012qf,Gao:2016aa} With the recent development of fluorescence detected nonlinear spectroscopy,\cite{Tekavec:2007aa,Yu:2019aa,Bruder:2019ua} double quantum coherence and multi-quantum resonances in general have been probed in atomic vapors~\cite{Bruder:2015vk,Yu:2019aa,Liang:2021uj} across a wide of range of densities including very low density vapors.\par
While it is well understood why the nonlinear DQS signal vanishes for uncorrelated electrons,\cite{Mukamel:2007aa} a complete theoretical description of two-dimensional DQS of atomic vapors and its precise link to interatomic interactions is still lacking. In addition, questions on the spectral lineshape of the two-dimensional DQS peaks and its link to homogeneous and inhomogeneous broadening have been raised. \cite{Tollerud:2016aa,Lomsadze:2018aa,Lomsadze:2020aa} In this context, the theoretical description introduced by Leegwater and Mukamel of atomic interactions in dense atomic vapors based on dipole-dipole interactions successfully described the effect of self-broadening in linear and nonlinear spectroscopy.\cite{Leegwater:1993aa,Leegwater:1994aa,Leegwater:1994ab} This atomistic model included explicitly inhomogeneous broadening from atomic interactions, the vectorial nature of the atomic transitions and also addressed some of the effect of nuclear motion. However nuclear motion was never treated explicitly in the simulation and it remained in the dense regime where Doppler broadening could be neglected. More recently, Ames and coworkers introduced a theoretical model to describe one-dimensional multi-quantum experiments of low density vapors.~\cite{Ames:2021vi,Ames:2022tx}  This model includes explicitly the far-field contribution to dipole-dipole interactions and Doppler broadening. However, still today the effect of Doppler broadening has not been explicitly described for two-dimensional DQS experiments and in particular for dense atomic vapors for which a transition between a regime dominated by inhomogeneous Doppler broadening and a regime dominated by homogeneous self-broadening can occur.\par
In this contribution, we use the model of Leegwater and Mukamel to describe the nonlinear spectroscopy of dense atomic vapors specifically focusing on DQS across the Doppler broadening regime and the self-broadening regime.  We include explicitly in our simulations the nuclear motion and Doppler broadening using the two-body approximation. One of the goals of this work is to provide a better understanding of the link between DQS signals and the interatomic interaction. In Sec.~\ref{sec:theory}, we resume the theoretical model of Leegwater and Mukamel\cite{Leegwater:1993aa,Leegwater:1994aa} that we will use to describe two-dimensional DQS. In Sec.~\ref{sec:results}, we present the numerical results which are then discussed based on the short-time approximation developed in Sec.~\ref{sec:twobody}. In Sec.~\ref{sec:doppler}, we include the effect of Doppler broadening and consider the case of a potassium vapor. Finally we conclude in Sec.~\ref{sec:conclusions}.
\section{Theoretical Model}
\label{sec:theory}
\subsection{Hamiltonian}
We consider an ensemble of $N$ neutral atoms in an atomic vapor interacting through dipole-dipole interactions.  Here we consider the case of s-p atomic transitions and assumed that each atom $i$ is described by a single ground state $\ket{g}_i$ corresponding to the angular momentum $J_g=0$ and a triply degenerated excited state $\ket{\alpha}_i$, $\alpha=x,y,z$ corresponding to the angular momentum $J_e=1$.  Therefore, we neglect the fine and hyperfine structure of the atomic state.  We also assume that the translation of the atoms can be treated completely classically and is decoupled from the atomic state, such that the coordinate of each atom $i$ will be denoted by $\vect{r}_i(t)$. In practice, we will neglect atomic collisions and assume simply uniform translation $\vect{r}_i(t) = \vect{r}_i(0) + \vect{v}_i t$. The electronic Hamiltonian is then written 
%
\begin{equation}
H_0(t) = \sum_{i=1}^N \sum_{\alpha=x,y,z} \omega_0 b_{i\alpha}^\dagger b_{i\alpha}\\  +\sum_{i\neq j} \sum_{\alpha,\beta} J_{\alpha\beta}(\vect{r}_{ij}(t)) b^\dagger_{i\alpha} b_{j\beta}
\label{eq:H0}
\end{equation}
%
where $\omega_0$ is the atomic transition frequency, where $b_{i\alpha}^\dagger$ and $b_{i\alpha}$ are the creation and annihilation operators corresponding to the s-p transition of each atom and defined by $b_{i\alpha} = \ket{g}_i\bra{\alpha}_i$ and where $\vect{r}_{ij}(t) = \vect{r}_i(t) - \vect{r}_j(t)$. In Eq.~(\ref{eq:H0}), we used a set of units such that $\hbar=1$ and the summations over $i$ and $j$ are performed over the $N$ atoms and the summations over $\alpha$ and $\beta$ are performed over the atomic electronic states $\alpha=x,y,z$. For dense vapors, the dipole-dipole interaction term is written
\begin{equation}
J_{\alpha\beta}(\vect{r}) = \frac{d^2}{4\pi \epsilon_0 r^3 } \left(  \delta_{\alpha,\beta}  - 3 \hat{r}_\alpha \hat{r}_\beta \right),
\label{eq:Jab}
\end{equation}
where $\hat{\vect{r}} = \vect{r}/r$ is a unitary vector, $d$ is the real valued atomic transition dipole moment, and $\epsilon_0$ is the vacuum permittivity. Note that  as the atomic coordinates $\vect{r}_i(t)$ depend explicitly on time, the Hamiltonian $H_0(t)$ is also explicitly time-dependent. 
\subsection{Linear response}
The interaction between the atomic vapor and the external electric field is written 
\begin{equation}
H_1(t)= - d \sum_{i\alpha}  \left( b_{i\alpha} + b_{i\alpha}^\dagger \right) E_\alpha (\vect{r}_i,t),
\end{equation}
where $E_{\alpha}(\vect{r}_i,t)$ is the component of the electric field in direction $\alpha$ at position $\vect{r}_i$ and time t. As described elsewhere, \cite{Berne:1970aa,Leegwater:1994aa,Schvaneveldt:1994aa} the linear susceptibility is written
\begin{equation}
\chi(\vect{k},\omega) = \sum_{\alpha }\frac{\ii}{3 \epsilon_0 V} \int_0^\infty \!\! \ud t \ e^{\ii \omega t} \big\langle  \left[  P_{\vect{k},\alpha}(t), P_{-\vect{k},\alpha}   \right] \big\rangle,
\end{equation}
where $V$ is the volume of the sample and where the polarization operator is defined by
\begin{equation}
P_{\vect{k},\alpha}(t) = d \sum_{i} \left( b_{i\alpha}(t) + b_{i\alpha}^\dagger(t)\right) e^{\ii \vect{k}\cdot  \vect{r}_i(t)}.
\label{eq:polarization}
\end{equation}
In Eq.~(\ref{eq:polarization}), the time evolution of the operators $b_{i\alpha}(t)$ and  $b^\dagger_{i\alpha}(t)$ arises from the Heisenberg representation of the creation and annihilation operators.  Here, we are considering a dense vapor and we will take the long wavelength limit $k\rightarrow0$,
\begin{equation}
\chi(\omega) = \frac{\ii d^2 n  }{ 3  \epsilon_0 } \int_0^\infty \!\!\! \ud t \ e^{\ii \omega t} \sum_\alpha \left( R_{\alpha\alpha}(t) - R_{\alpha\alpha}^*(t) \right),
\end{equation}
where $n=N/V$ is the number density and where the linear response function $R_{\alpha\alpha}(t)$ is given by 
\begin{equation}
R_{\alpha\alpha}(t) = \frac{1}{N} \bra{g}D_\alpha U(t,0) D^\dagger_{\alpha} \ket{g}.
\label{eq:linearresponse}
\end{equation}
In Eq.~(\ref{eq:linearresponse}), we have introduced the system ground state $\ket{g} = \prod_{i=1}^{N} \ket{g}_i$. The time evolution operator is defined by the time ordered exponential
\begin{equation}
U(t_2,t_1) = \exp_+\left( -\ii \int_{t_1}^{t_2} \ud \tau H_0(\tau) \right),
\end{equation}  
and the lowering and raising operators $D_\alpha$ are defined by
\begin{equation}
D_\alpha^{\dagger} = \sum_{i} b^\dagger_{i \alpha}.
\end{equation}
Finally, using the fact that $\omega_0$ is much larger than the spectral linewidth, the absorption spectrum is computed using 
\begin{equation}
I(\omega) = \im \chi(\omega)  \approx \frac{d^2 n }{3 \epsilon_0 } \sum_{\alpha} \re \int_0^\infty \!\!\! \ud t \ e^{\ii \omega t} R_{\alpha\alpha}(t).
\end{equation}
\subsection{Fluorescence detected double quantum spectroscopy}
In a fluorescence detected double quantum spectroscopy (FDQ) experiment, the atomic system interacts with a set of four laser pulses to create a fourth order population which then relaxes through spontaneous emission. We will assume that the duration of each pulse is short compared to the time evolution of the system and that pulse $p$ is injected in the system at time $\tau_p$. To simplify, we will fix the origin of time such that $\tau_1=0$, and if we denote $t_1$, $t_2$ and $t_3$ the time intervals between the pulses, we have $\tau_2=t_1$, $\tau_3=t_1+t_2$ and $\tau_4=t_1+t_2+t_3$. Denoting  $\phi_p$ and $\bm{\epsilon}^{(p)}$ the phase and polarization of pulse $p$,  The total external electric field of the pulses is written
%
\begin{equation}
\vect{E}(\vect{r},t) = \sum_{p=1}^4 \bm{\epsilon}^{(p)}  \mathcal{E}_L(t-\tau_p) e^{\ii \vect{k}_L\cdot \vect{r} - \ii \omega_L (t-\tau_p) + \ii \phi_p } \\ + \text{h.c.},
\end{equation}
%
where $\vect{k}_L$, $\omega_L$ and $\mathcal{E}_L(t)$ are the wave vector, central frequency and  enveloppe respectively. Here we will only consider a configuration where all pulses have the same wavevector chosen along  the  $z$ direction of the laboratory frame $\vect{k}_L = k_0 \vect{e}_z$ and have the same polarization chosen along the $x$ direction  $\bm{\epsilon}^{(p)} = \vect{e}_x$. We can easily generalize for other polarization configurations. Using the rotating wave approximation, and taking the long wavelength limit,  the interaction between the atomic system and the laser pulses can be written
\begin{equation}
H_1 =  - D_x^\dagger E^{-}(t) -  D_x  E^{+}(t),
\end{equation}
where $E^{+}(t) = \left(E^{-}(t) \right)^* $ with $E^{-}(t)$  defined by
\begin{align}
E^{-}(t) =  \sum_p \mathcal{E}_L(t-\tau_p) e^{- \ii \omega_L (t-\tau_p) + \ii \phi_p }.
\end{align}
Assuming that the system completely relaxes and that all photons are collected, the FDQ signal is then given by\cite{Perdomo-Ortiz:2012aa,Schroter:2018aa} 
\begin{equation}
S_{\text{DQ}} = \tr \left\{ P \rho^{(4)} (\tau_4) \right\},
\label{eq:relaxation}
\end{equation}
where $P= \sum_{i\alpha} b^\dagger_{i\alpha} b_{i\alpha}$ is the population operator and where $\rho^{(4)}(\tau_4)$ is the fourth order population generated by the four laser pulses. Using Eq.~(\ref{eq:relaxation}), we have also assumed that there is no non-radiative decay processes in the atomic system. However, at high density, molecular formation between alkali atoms can occur.\cite{Vdovic:2006aa} The molecular absorption resulting from diatomic molecules is shifted and do not interfere with the atomic D$_1$ and D$_2$ lines that we are presently modeling. However, during the fluorescence decay a small fraction of atoms will likely go through molecular formation opening a non-radiative decay channel. If we assume that the overall fluorescence quantum yield remains proportional to the number of excited atoms in the system then the DQS signal is again given by Eq.~(\ref{eq:relaxation}) but with an additional factor of proportionality corresponding to an atomic fluorescence quantum yield lower than one. As a result, we do not expect a large difference in our results if we include molecular formation and we will completely neglect it at this point.\par
 The interaction between the atomic system and the laser pulses induce a set of Liouville pathways. With a conventional two-dimensional spectroscopy experiment, a subset of pathways is selected using the phase-matching condition resulting in different experimental measurements such as rephasing ($\vect{k}_{\text{I}}$), nonrephasing ($\vect{k}_{\text{II}}$) and double-quantum spectroscopy ($\vect{k}_{\text{III}}$). In a fluorescence detected experiment, the phase-matching condition is obtained through cycling of the phase $\phi_p$ in order to obtain the phase-matching corresponding to the nonlinear technique we are considering. In the case of a DQS experiment we choose a global phase $\phi_1 + \phi_2 - \phi_3 -\phi_4$. Using perturbation theory,  the fourth order population corresponding to that global phase is given by
\begin{equation}
\rho^{(4)}(\tau_4) = \ii d \mathcal{E}_L \tau_L  \mathscr{D}\rho^{(3)}(\tau_4),
\end{equation}
where $\mathscr{D}= \left[ D_x,\cdot \right]$ is the Liouville operator associated with the lowering operator $D_x$, $\tau_L$ is the pulse duration and where $\rho^{(3)}(\tau_4)$ is the third-order polarization generated by the first three pulses and is written as
%
\begin{equation}
\rho^{(3)}(\tau_4) =\left( \ii d  \mathcal{E}_L \tau_L\right)^3 \mathscr{U}(\tau_4,\tau_3)  \mathscr{D}  \mathscr{U}(\tau_3,\tau_2) \mathscr{D}^\dagger  \mathscr{U}(\tau_2,\tau_1) \mathscr{D}^\dagger \rho_0. 
\label{eq:rho3}
\end{equation}
%
In Eq.~(\ref{eq:rho3}), $\rho_0$ is the initial density matrix and $\mathscr{U}(\tau_2,\tau_1)$ is the evolution operator in Liouville space which is expressed as a function of the Liouville operator $\mathscr{L}_0(t) = \left[ H_0(t), \cdot \right]$ 
\begin{equation}
\mathscr{U}(\tau_2,\tau_1) = \exp_+\left(- \ii \int_{\tau_1}^{\tau_2} \! \! \! \mathscr{L}_0(\tau) \ud \tau\right).
\end{equation}
Here we assume that the atomic system is initially in the electronic ground states $\rho_0=\ket{g}\bra{g}$.\par
As the lowering operator $D_x$ destroys an excitation we have the commutation relation
\begin{equation}
\left[ D_x, P \right] = D_x P - PD_x = D_x,
\end{equation}
which gives the expression of the DQS signal as
\begin{equation}
S_{\text{DQ}} \propto \ii \tr\left\{ P \left[ D_x,\rho^{(3)}(\tau_4) \right] \right\} \\
 =  - \ii \tr\left\{D_x \rho^{(3)}(\tau_4) \right\}  = - R^{\vect{k}_{\text{III}}}_{xxxx}.
\end{equation}
Therefore we recover the well-known result that the fluorescence detected signal is, up to a minus sign, simply given by the polarization detected double quantum signal.\cite{Perdomo-Ortiz:2012aa}  This directly results from the fact that we have neglected any non-radiative decay processes and that the fluorescence quantum yield is simply proportional to the number of excited atoms in the system. Using the notation from Ref.~\citenum{Perdomo-Ortiz:2012aa}, this corresponds to a value of the parameter $\Gamma=2$ which gives the number of photons emitted from the two-exciton states. It is easy to show that the DQS signal is  composed of two Liouville pathways\cite{Mukamel:2004fk}
\begin{equation}
R^{\vect{k}_{\text{III}}}_{xxxx} = R^{(3a)}_{xxxx} - R^{(3b)}_{xxxx},
\end{equation}
where the correlation functions $R^{(3a)}_{xxxx}$ and $R^{(3b)}_{xxxx}$ are given, for example, in Refs.~\citenum{Mukamel:2004fk} and~\citenum{Abramavicius:2009fk}.
To simplify, we are going to consider only the case where the two first pulses are simultaneous ${t_1=\tau_2-\tau_1=0}$. It is easy to show that the correlation functions $R^{(3a)}_{xxxx}$ and $R^{(3b)}_{xxxx}$ are written
\begin{align}
&R^{(3a)}_{xxxx}(t_2,t_3) =\bra{g} D_x U(t_2+t_3,t_2) D_x  U(t_2,0) D^\dagger_x  D^\dagger_x \ket{g} \label{eq:R3a}, \\ 
&R^{(3b)}_{xxxx}(t_2,t_3) =   \bra{g}   D_x U(t_2,t_2+t_3 ) D_x U(t_2+t_3,0) D^\dagger_x  D^\dagger_x \ket{g} .\label{eq:R3b} 
\end{align}
The signal in the frequency domain is then given by
\begin{equation}
S_{\text{DQ}}(\omega_2,\omega_3) = \int_0^\infty \!\!\! \int_0^\infty \!\!\! \ud t_2 \ud t_3 \ e^{\ii \omega_2 t_2 + \ii \omega_3 t_3 } S_{\text{DQ}}(t_2,t_3).
\end{equation}
\subsection{Reduced units}
As pointed out in Ref.~\citenum{Leegwater:1994aa}, for a vapor at number density $n=N/V$, the average distance between neighboring atoms allows us to both define a unit of length $r_0$ and a unit of energy $E_0$ through the average dipole-dipole interaction. For a dilute gas, the probability distribution for the vector $\vect{r}$ between two nearest neighboring atoms is given by\cite{Torquato:1995aa}
\begin{equation}
P(\vect{r}) \ud^3 \vect{r}  = n r^2  e^{-\frac{4\pi}{3} n r^3} \ud r \ud \Omega,
\end{equation}
where $r$ is the nearest neighbor distance and $\ud \Omega$ is the infinitesimal element of solid angle. 
The average distance between two neighboring atoms is then given by
\begin{equation}
\overline{r} = \int   r\  P(\vect{r}) \ud^3 \vect{r}= \left( \frac{4\pi n}{3}\right)^{-\frac{1}{3}} \Gamma\left( \frac{4}{3}\right).
\end{equation}
with $\Gamma(4/3)\approx 0.89298$. Therefore, following Ref.~\citenum{Leegwater:1994aa}, we define the unit of length as
\begin{equation}
r_0  =  \left( \frac{4\pi n}{3}\right)^{-\frac{1}{3}},
\end{equation}
and the unit of energy $E_0$ corresponding to the strength of the dipole-dipole interaction at  distance $r_0$
\begin{align}
E_0 = \frac{d^2}{4\pi \epsilon_0 r_0^3} = \frac{n d^2}{3\epsilon_0} = n \frac{\Gamma_0\pi c^3}{\omega_0^3},
\end{align}
where $\Gamma_0$ is the natural linewidth. As pointed out in Ref.~\citenum{Leegwater:1994aa}, for densities higher than $10^{15}$, the average distance between atoms is much smaller than the optical wavelength but much larger than the range of the core potential. In this case, there is only one energy scale given by $E_0$. Using the reduced units $r_0$ and $E_0$ allows us to completely eliminate the density $n$ from the problem.\par
\begin{table}
\begin{tabular*}{\linewidth}{@{\extracolsep{\fill}}ll }
\hline\hline
 $\omega_0 /2\pi$ (THz) &  390.5 \\
 $m$ (amu) & 39.0983 \\
$\Gamma_0$ (ps$^{-1}$) & 3.76$\times10^{-5}$  \\
\hline\hline
\end{tabular*}
\caption{Values of the model parameters for a potassium vapor}
\label{tab:potassium}
\end{table}
\begin{figure}
\centering
\includegraphics{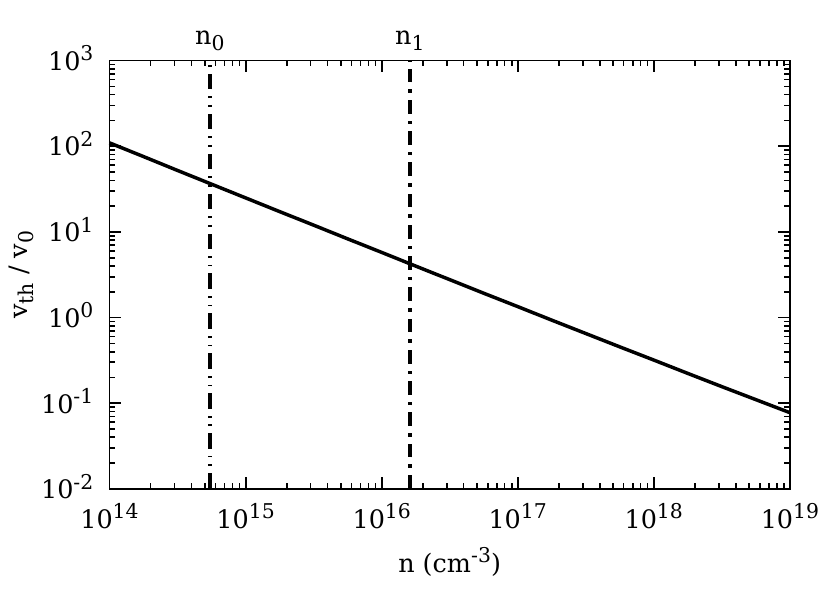}
\caption{Parameter $v_{\textrm{th}}/v_0$ as a function of density for a potassium vapor}
\label{fig:vthv0}
\end{figure}
The typical time required for an electronic excitation to delocalize from one atom to its nearest-neighbor is given by $t_0=1/E_0$.  The nuclear motion will only have an important role if the distance traveled by the atoms during this delocalization time is large or comparable to the average distance between neighboring atoms. Therefore, we introduce the unit of velocity as
\begin{equation}
v_0 = r_0 E_0 =   \left( \frac{4\pi }{3}\right)^{-\frac{1}{3}} \frac{n^{2/3} d^2}{3 \epsilon_0}.
\end{equation}
If we assume that the atoms are at thermal equilibrium at temperature $T$ with a thermal velocity
\begin{equation}
v_{\text{th}} = \sqrt{k_B T/m},
\end{equation}
 where $m$ is the atomic mass, the important parameter that controls the effect of nuclear motion is therefore $v_{\text{th}}/v_0$. In Fig.~\ref{fig:vthv0}, we represent the evolution of the parameter  $v_{\text{th}}/v_0$ as a function of the density in the case of a potassium vapor where we have used the expression of the density as a function of temperature $n(T)$ from Ref.~\citenum{Nesmeyanov:1963aa}. The parameters for potassium are obtained by averaging the properties of the $^2P_{1/2}$ and $^2P_{3/2}$ states as given by the NIST atomic spectra database.\cite{NIST_ASD} The final parameters are resumed in Table~\ref{tab:potassium}. In Fig.~\ref{fig:vthv0}, we also report the value of the density 
 \begin{equation}
 n_0 = k_0^3 = \left(\omega_0 /c\right)^3 \approx 5.5 \times 10^{14}~\text{cm}^{-3},
 \end{equation}
 under which the long wavelength limit is no longer valid. Therefore, densities $n\lesssim n_0$ will not be properly described by the present theoretical model.
We also show in Fig.~\ref{fig:vthv0} the value of the density $n_1$ for which self broadening and Doppler mechanisms are of the same order. The value of $n_1$ is estimated by assuming that Doppler broadening is given by 
 \begin{equation}
 \Gamma_{\text{D}} = 2 \sqrt{2 \log 2} \omega_0 v_\text{th} / c
 \end{equation}
 and self-broadening by 
 \begin{equation}
\Gamma_{\text{self}} = 2E_0
\end{equation}
 and by solving $\Gamma_{\text{D}} = \Gamma_{\text{self}}$ for $n_1$. For a potassium vapor we found $n_1 \approx  1.6 \times 10^{16}~\text{cm}^{-3}$. In this article, we will consider vapor densities across the self-broadening and Doppler regimes such that $n \gg n_0$. Fig.~\ref{fig:vthv0} shows that the nuclear motion will only have an important impact at a low density close or lower than $n_1$ mostly because of a slower electronic dynamics. As decreasing the density we increase slightly the temperature and the thermal velocity however the main effect arises from the fact that as the average distance between atoms increases, the interaction energy decreases and the typical time necessary for the wavefunction to delocalize increases. Therefore, the wavefunction is more prone to be perturbed by nuclear motion in this regime. In the case of a potassium vapor, at the density $n=n_1$ we have $v_{\text{th}}/v_0\approx 4.2$.
\subsection{Numerical propagation}
Without loss of generality, we set in our simulations $\omega_0=0$ and we perform a direct numerical integration of the wavefunction in reduced units to obtain the linear and nonlinear responses (Eqs.~\ref{eq:linearresponse},~\ref{eq:R3a}, and ~\ref{eq:R3b}). To simulate the atomic vapor, we use periodic boundary conditions and introduce an ensemble of $N=48$ atoms in a cubic simulation box of length $L$. In reduced units, the box length is given by 
\begin{equation}
L=  \left(3N/4\pi\right)^{1/3}.
\end{equation}
At time $t=0$, the coordinates of the atoms are taken randomly within the volume of the box and their velocities are taken randomly using the Maxwell-Boltzmann distribution at temperature $T$. The linear and nonlinear signals are averaged over a set of $10^5$ initial conditions. To improve sampling, we also perform orientational averaging of the nonlinear signal as described in Ref.~\citenum{Andrews:1977fk}. The third-order response function where all pulse polarization are aligned in the same direction is then given by
 \begin{equation}
 \overline{R}^{\vect{k}_{\text{III}}}_{xxxx} = \frac{1}{15} \sum_{\alpha,\beta} \left( R^{\vect{k}_{\text{III}}}_{\alpha\alpha\beta\beta}  + R^{\vect{k}_{\text{III}}}_{\alpha\beta\alpha\beta} + R^{\vect{k}_{\text{III}}}_{\alpha\beta\beta\alpha} \right).
 \end{equation}
 As in Ref.~\citenum{Leegwater:1994aa} we also include explicitly  long-range interactions using the Ewald summation. We simply modify the coupling constants $J_{\alpha\beta}(\vect{r}_{nm})$ in the expression of $H_0$ by including explicitly periodic boundary conditions
\begin{equation}
J_{\alpha\beta}(\vect{r}_{ij}) \rightarrow \sum_{\vect{p}\in \mathbb{Z}^3} J_{\alpha\beta}(\vect{r}_{ij} + \vect{p} L ).
\end{equation}
In practice we use the expressions derived from the Ewald methods as described in Ref.~\citenum{Leeuw:1980aa} and applied here for the calculation of $J_{\alpha\beta}(\vect{r}_{ij})$.\par
The tensor $J_{\alpha\beta}(\vect{r}_{ij})$ decays with the interatomic distance as $r_{ij}^{-3}$ and can increase very drastically as the distance between the atom decreases.  Therefore, a traditional Runge-Kutta integrator with a fixed timestep  will fail unless one takes a ridiculously small timestep.  In order to fasten the integration process we assume that the wavefunction will mostly delocalize among nearest neighbors. Indeed, for the densities we consider, the probability to have three atoms or more in a given neighborhood (e.g. in a sphere of radius $R$) is  negligibly small compared to the probability to have two atoms. Therefore, we will perform the following approximations. First, we assume that over a small timestep $\Delta t$, the atoms do not move significantly such that the time dependence of the tensor $J_{\alpha\beta}(\vect{r}_{ij})$  can be neglected. Second, over this time step $\Delta t$, the time-evolution operator can be factorized as a product of two-body terms.%
\begin{equation}
\ket{\psi(t+\Delta t)} = \prod_{i<j} e^{-\ii V_{ij}(t) \Delta t}\ket{\psi(t)} + \mathcal{O}(\Delta t^2),
\label{eq:propagate1dense}
\end{equation}
where $V_{ij}(t)$ is a $6\times6$ matrix defined by 
\begin{equation}
V_{ij}(t)= \sum_{\alpha\beta} J_{\alpha\beta}\left(\vect{r}_{ij}(t)\right)  \left( b^\dagger_{i\alpha} b_{j\beta} + b_{i\alpha} b^\dagger_{j\beta}\right).
\end{equation}
Note that Eq.~(\ref{eq:propagate1dense}) is exact in the limit $\Delta t\rightarrow 0$. As we use the Ewald summation, the matrix $V_{ij}(t)$ cannot be diagonalized analytically but can be written in a block diagonal form. Using the basis 
\begin{equation}
\ket{\pm,\alpha}_{ij} = \frac{1}{\sqrt{2}} \left( b^\dagger_{i\alpha} \pm b^\dagger_{j\alpha}  \right) \ket{g},
\label{eq:basismol}
\end{equation}
the two-body matrix $V_{ij}$ can be written 
\begin{equation}
V_{ij}^{\alpha\beta} = \begin{pmatrix} J_{\alpha\beta}(\vect{r}_{ij})  &0  \\  0 & -J_{\alpha\beta}(\vect{r}_{ij}) \end{pmatrix}.
\end{equation}
If we denote $J_{\lambda}$ and $\phi_{\lambda \alpha}$ the real-valued  eigenvalues and eigenstates  of the $3\times3$ matrix  $J_{\alpha\beta}(\vect{r}_{ij})$
\begin{equation}
\sum_\beta J_{\alpha\beta} \phi_{\lambda \beta} = J_{\lambda}\phi_{\lambda \alpha},
\end{equation}
we can then write the two-body propagator as
\begin{equation}
e^{-\ii V_{ij}\Delta t} = \sum_\lambda \sum_{\alpha\beta}  \phi_{\alpha \lambda} \phi_{\beta \lambda} \left[ \cos\left( J_\lambda \Delta t\right)  \left( b_{i\alpha}^\dagger b_{i\beta} +  b_{j\alpha}^\dagger b_{j\beta} \right)  - \ii \sin\left( J_\lambda \Delta t\right)  \left( b_{i\alpha}^\dagger b_{j\beta} +  b_{j\alpha}^\dagger b_{i\beta} \right) \right]. 
\label{eq:propagator}
\end{equation}
\par
In practice, at each time step, we diagonalize numerically the $3\times3$ coupling matrix $J_{\alpha\beta}(\vect{r}_{ij})$ that includes the Ewald summation using the efficient algorithm described in Ref.~\citenum{Kopp:2008we} and we propagate the one-exciton and two-exciton wavefunctions using Eqs.~(\ref{eq:propagate1dense}) and~(\ref{eq:propagator}). These wavefunctions are then used to compute the correlation functions given by Eqs.~(\ref{eq:R3a}) and~(\ref{eq:R3b}).\par

 During the simulations, we used a value of the time step $\Delta t$ that depends on the thermal velocity. We used $\Delta t= \pi/100$ for $v_{\text{th}}=0$ and $0.1$, $\Delta t=\pi/150$ for $v_{\text{th}}=0.2$,   $\Delta t=\pi/200$ for $v_{\text{th}}=0.5$, $1.0$ and $2.0$,  $\Delta t=\pi/400$ for $v_{\text{th}}=5.0$ and  $\Delta t=\pi/1200$ for $v_{\text{th}}=10.0$.
\section{Numerical Results}
\label{sec:results}
\subsection{Linear absorption}
\begin{figure}
\centering
\includegraphics{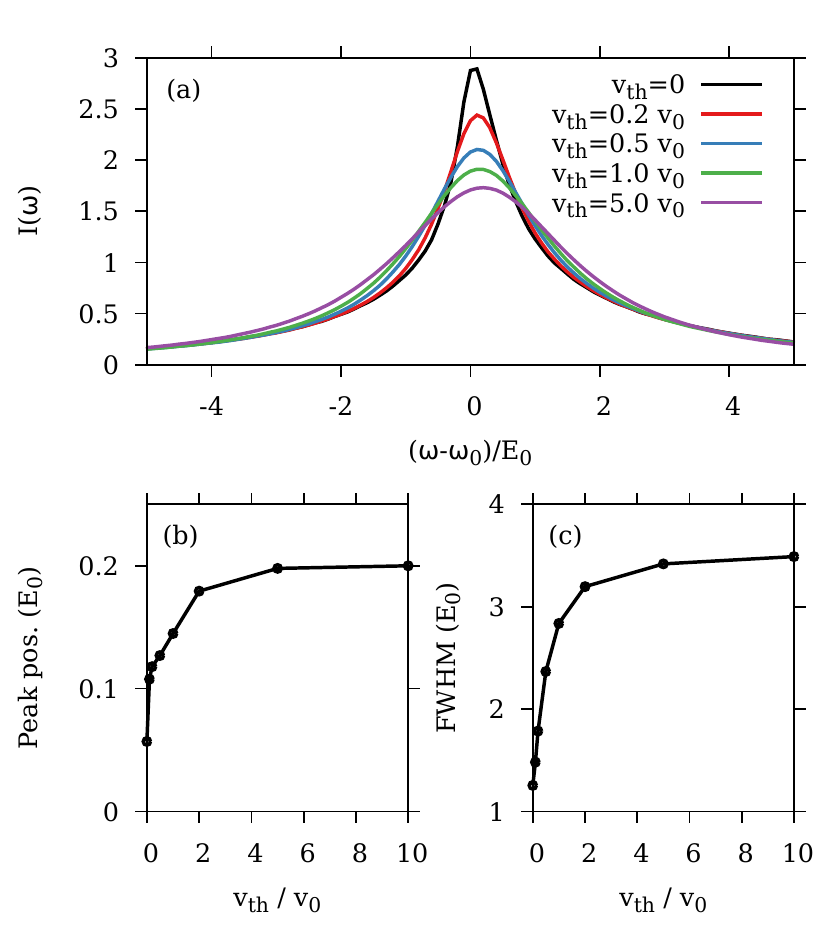}
\caption{(a) Linear absorption spectrum $I(\omega)$ as a function of the frequency shift $(\omega-\omega_0)/E_0$ and for different values of $v_{\text{th}}$. (b) Peak position of the linear absorption spectrum as a function of the thermal velocity $v_{\text{th}}$. (c) Full-width-half-maximum of the linear absorption spectrum as a function of the thermal velocity $v_{\text{th}}$.}
\label{fig:linear}
\end{figure}
Fig.~\ref{fig:linear}a shows the linear absorption spectrum $I(\omega)$ as a function of the frequency shift $(\omega-\omega_0)$ and for different values of the thermal velocity $v_{\text{th}}$. In the static case $v_{\text{th}}=0$, we recover the non-Lorentzian asymmetric lineshape obtained in Ref.~\citenum{Leegwater:1994aa}. As increasing the thermal velocity, the linear absorption spectrum undergoes both a broadening of the lineshape and a shift of its peak position  towards the higher frequencies keeping the integrated spectrum constant. Note that as the thermal velocity is increased the spectrum loses its asymmetric lineshape to adopt a more Lorentzian lineshape. To quantify this behavior, we represent in Figs.~\ref{fig:linear}b and~\ref{fig:linear}c the peak position and full-width-half-maximum (FWHM) of the absorption spectrum as a function of the thermal velocity $v_{\text{th}}$ respectively. In the static case, the peak position and the FWHM are given by $0.0591$ and $1.25$, respectively. As the thermal velocity increases both the peak position and FWHM increase first linearly and then reaches a maximum value for large values of the thermal velocity. This confirms the findings of Ref.~\citenum{Leegwater:1994aa} that collisional broadening is completely independent of  temperature in the fast collision regime.  For $v_{\text{th}} =10$, the values of the peak position and FWHM are given by $0.2$ and $3.49$ respectively. Note that the value of the FWHM is in between the two values of $2.21$ and $5.76$ proposed in Ref.~\citenum{Leegwater:1994aa} depending on the model and where the collisional broadening was included using a semi-classical approach.
\subsection{Double quantum spectroscopy}
\begin{figure}
\centering
\includegraphics{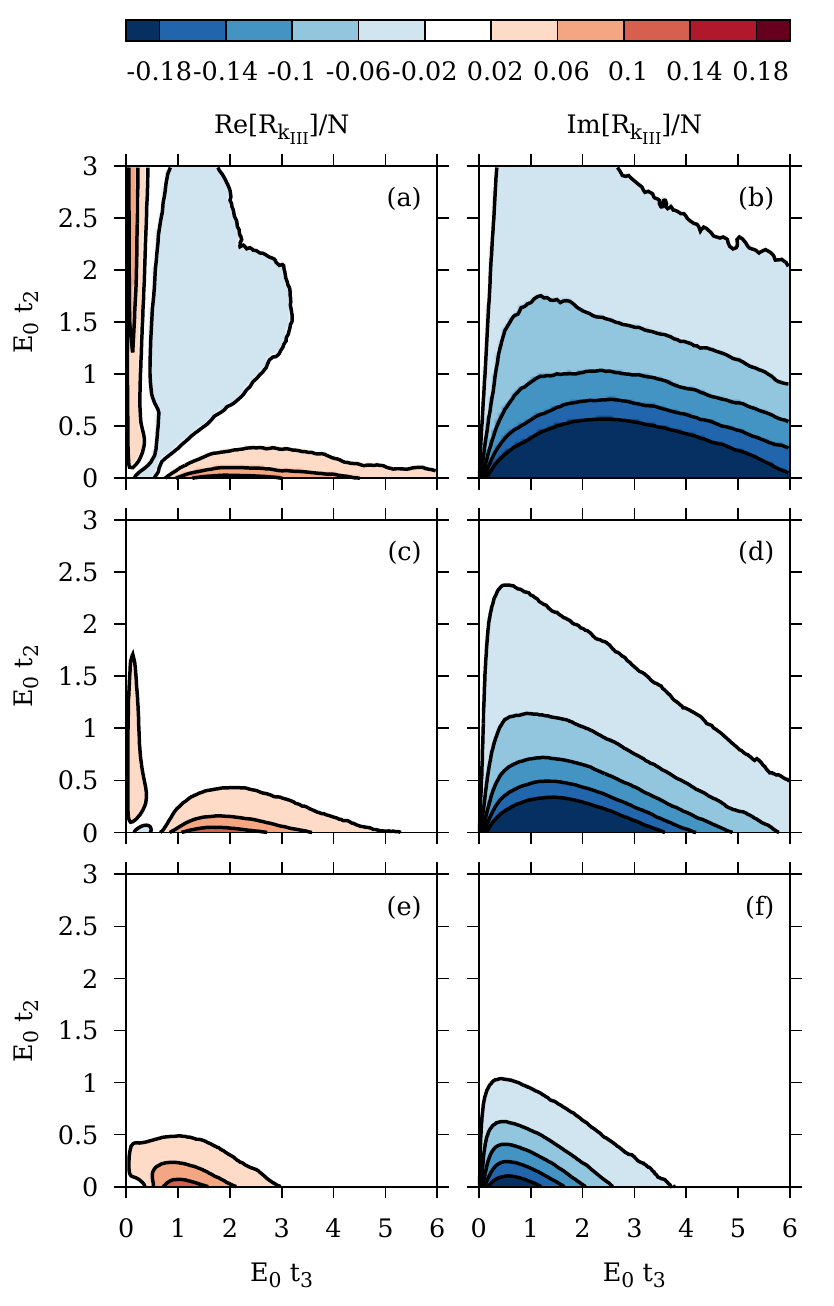}
\caption{Real part (left panel) and imaginary part (right panel) of the double quantum spectrum per atom in the time domain $R_{\vect{k}_{\text{III}}}(t_2,t_3)/N$ for  (a,b) $v_{\text{th}}=0$,  (c,d)  $v_{\text{th}}=0.2 v_0 $, and  (e,f)  $v_{\text{th}}=v_0$.}
\label{fig:k3time2d}
\end{figure}
\begin{figure}
\centering
\includegraphics{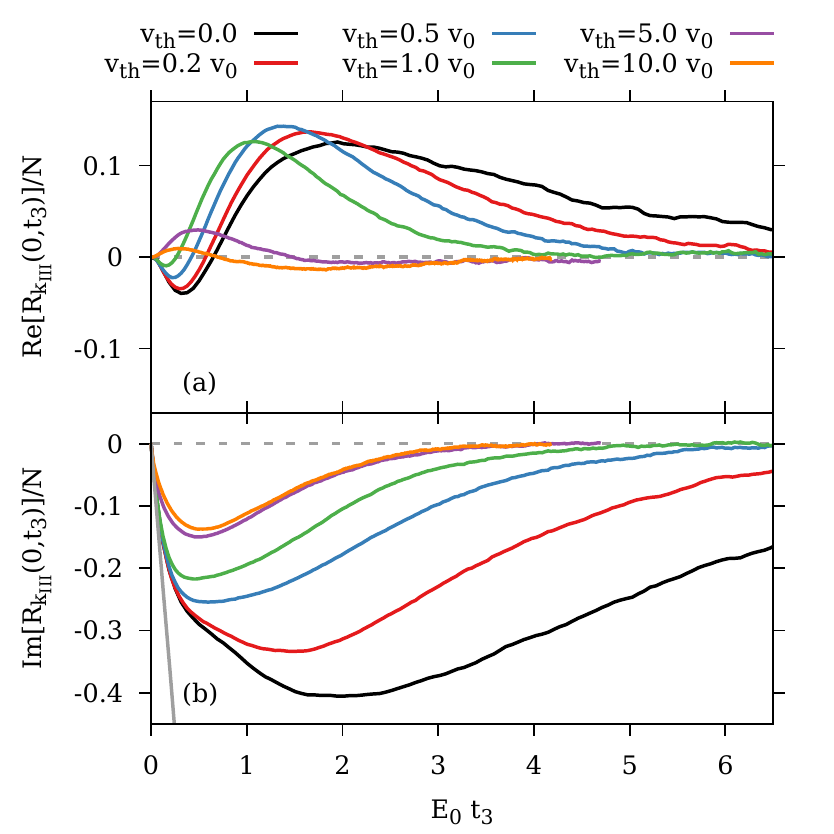}
\caption{Real part (a) and imaginary part (b) of the double quantum spectrum per atom in the time domain for $t_2=0$ $R_{\vect{k}_{\text{III}}}(0,t_3)/N$ and for different values of the thermal velocity $v_{\text{th}}$.}
\label{fig:k3timet3}
\end{figure}

Fig.~\ref{fig:k3time2d} shows the real and imaginary part of the DQS signal per atom $R_{\vect{k}_{\text{III}}}(t_2,t_3)/N$ in the time domain as a function of the times $t_2$ and $t_3$ and for different values of the thermal velocity $v_{\text{th}}$. In the static case $v_{\text{th}}=0$ (Figs.~\ref{fig:k3time2d}a and \ref{fig:k3time2d}b), the DQS signal is dominated by its imaginary part and its absolute value shows a maximum located at $t_2=0$ and $t_3 = 1.9$ with a typical decay time of $5.6$ along the axis $t_3$ and a much faster decay time of $0.44$ along the axis $t_2$. This is more clearly seen in Fig.~\ref{fig:k3timet3} which shows the nonlinear signals as a function of $t_3$ and for $t_2=0$. In Fig.~\ref{fig:k3time2d}a, we also notice an interesting but small feature in the real part of the DQS signal  for non vanishing but small values of $t_3$ with a very slow decay along $t_2$ of about $9.7$. Figs.~\ref{fig:k3time2d} and~\ref{fig:k3timet3} show that, as increasing the thermal velocity the overall spectrum decays faster and the real part of the spectrum vanishes, including the small feature described earlier. This is clearly seen in Fig.~\ref{fig:k3timet3}, which shows that for large values of $v_{\text{th}}$, the imaginary part of the signal  as a function of $t_3$ becomes bell shaped with a linear increase for small values of $t_3$ and an exponential decay for large values of $t_3$. For $v_{\text{th}}=10 v_0$, the real part of the signal has almost completely vanished and  the typical decay time of the nonlinear signal is given by $0.2$ along $t_2$ and $1.6$ along $t_3$.\par
\begin{figure}
\centering
\includegraphics{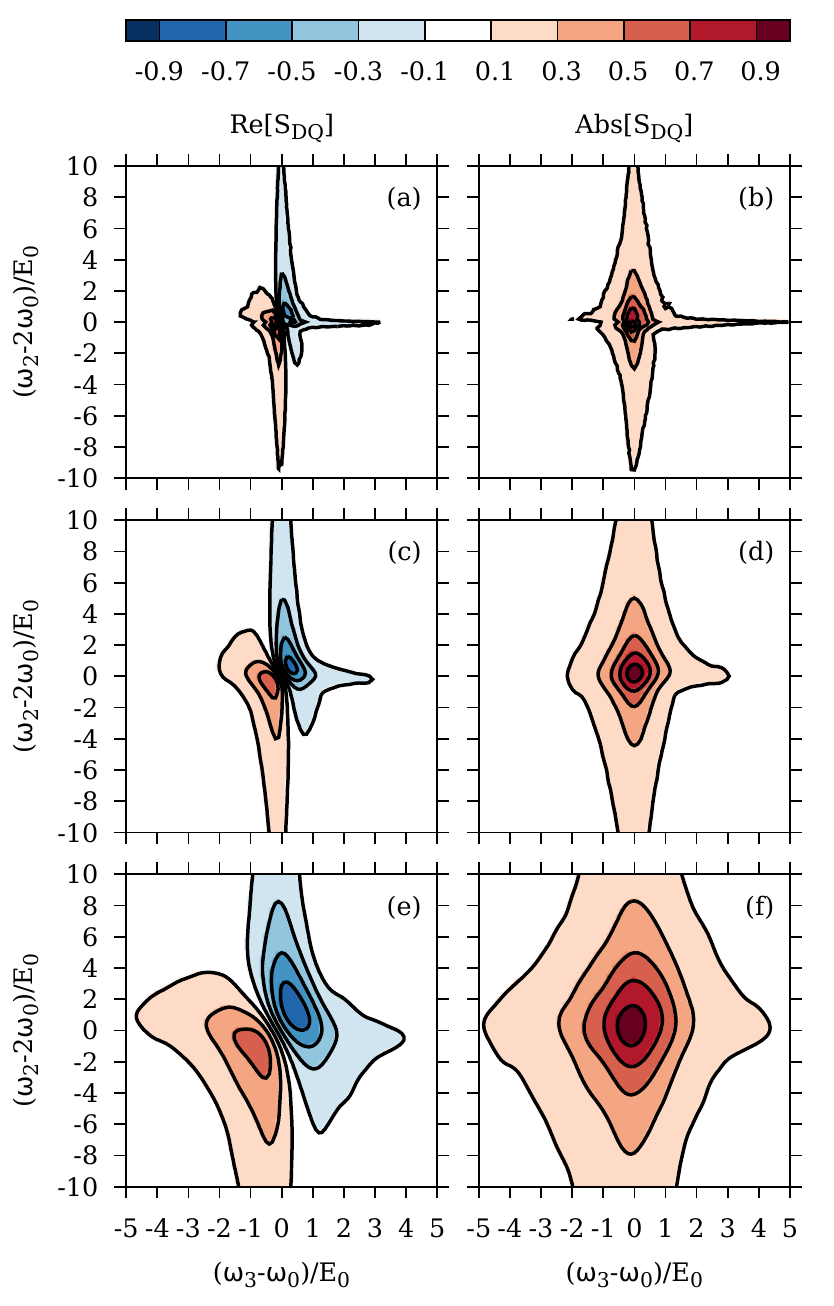}
\caption{Real part (left panel) and absolute value (right panel) of the normalized double quantum spectrum in the frequency domain $S_{\text{DQ}}(\omega_2,\omega_3)$ for  (a,b) $v_{\text{th}}=0$,  (c,d)  $v_{\text{th}}=0.2 v_0 $, and  (e,f)  $v_{\text{th}}=v_0$.}
\label{fig:k3freq2d}
\end{figure}
While fluorescence detected nonlinear signals are naturally measured in the time domain, they are often represented in the frequency domain in order to highlight resonances. Fig.~\ref{fig:k3freq2d} shows the real part of the DQS signal  $S_{\text{DQ}}(\omega_2,\omega_3)$ in the frequency domain as a function of the frequency shifts $(\omega_2-2\omega_0)$ and $(\omega_3-\omega_0)$ and for different values of the thermal velocity $v_{\text{th}}$. We also represent the absolute value of the different spectra as it is often used in experimental studies. In order to compare the nonlinear spectra for different values of the thermal velocity, in Fig.~\ref{fig:k3freq2d} the double quantum spectra are normalized with respect to the maximum of the absolute value. As noticed in Ref.~\citenum{Dai:2012qf}, the real part of the DQS signal shows a dispersive profile corresponding to the first derivative of an absorption peak. This can be directly explained by the linear behavior of the imaginary part of the response function for small values of $t_3$ as observed earlier in Fig.~\ref{fig:k3timet3}. Indeed, the derivative of an absorption peak in the frequency domain with respect to $\omega_3$ will result in the multiplication by a factor $\ii t_3$ in the time domain. 
For $v_{\text{th}}=0$, the linewidths of the DQS signal, corresponding to the FWHM of the absolute value are $3.1$ along the axis $\omega_2$ and $0.85$ along the axis $\omega_3$. Fig.~\ref{fig:k3freq2d} shows that, an increasing value of the thermal velocity induces a strong broadening of the spectrum. This is a direct consequence of the faster decay observed in the time domain. As previously, for large values of the thermal velocity the DQS signal becomes constants and for  $v_{\text{th}}=10v_0$ (data not shown), the linewidths are given by  $11.4$ along the axis $\omega_2$ and $3.4$ along the axis $\omega_3$. We also notice that the absolute value of the spectrum does not show any tilted or elongated shape as it has been observed experimentally for some systems.\cite{Tollerud:2016aa,Lomsadze:2018aa,Lomsadze:2020aa}\par
 The thermal velocity induces not only a broadening of the spectrum but also induces a strong decrease of the overall spectrum amplitude as can be seen in the time domain in Fig.~\ref{fig:k3timet3}.  To quantify this, we need to define a measure of the overall spectrum amplitude. The DQS signal in the time domain vanishes for $t_3=0$ since the two Liouville pathways defined in Eqs.~(\ref{eq:R3a}) and~(\ref{eq:R3b}) cancel each other. Consequently, the integral of the DQS signal in the frequency domain will also vanish. Therefore, we define the total spectrum amplitude as the integral of the absolute value of the spectrum in the frequency domain
 \begin{equation}
 A_{\text{DQ}} = \int_{2\omega_0-2\Delta \omega}^{2\omega_0+2\Delta\omega} \int_{\omega_0-\Delta\omega}^{\omega_0+\Delta\omega} \!\!\!  \ud \omega_2 \ud \omega_3   \left|S_{\text{DQ}}(\omega_2,\omega_3)\right|,
  \label{eq:IDQ}
  \end{equation}
where $\Delta\omega$ is a parameter which defines the integration box. Note that as shown in the Appendix, $\Delta\omega$ needs to remain finite since in the case of Lorentzian lineshapes, the integration over $\omega_2$ diverges for $\Delta\omega\rightarrow\infty$. Here we use the value $\Delta \omega = 50 E_0$. With this definition,  $A_{\text{DQ}}$ is dimension-less  and therefore  is completely independent of the energy scale $E_0$.
 \begin{figure}
\centering
\includegraphics{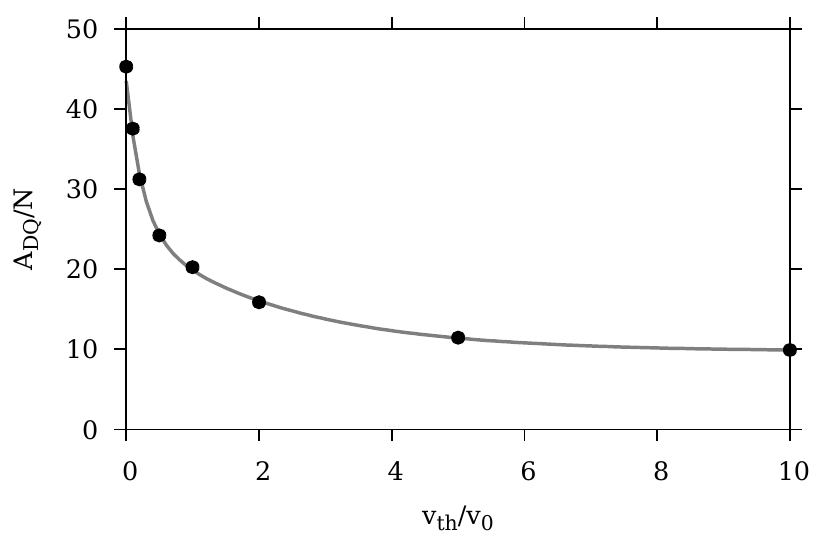}
\caption{Total double quantum spectrum amplitude $A_{\text{DQ}}$ as a function of the thermal velocity $v_{\text{th}}$. The gray line is added as a guide to the eye. }
\label{fig:k3total}
\end{figure}
 Fig.~\ref{fig:k3total} shows the dependence of the total spectrum amplitude $A_{\text{DQ}}$ defined by Eq.~(\ref{eq:IDQ}) as a function of the thermal velocity with $\Delta\omega=50 E_0$.  $A_{\text{DQ}}$ experiences a fast decay with the thermal velocity and eventually becomes constant  for large values of $v_{\text{th}}$, with an overall amplitude more than four times smaller than its value in the static case.\par
 \begin{figure}
\centering
\includegraphics{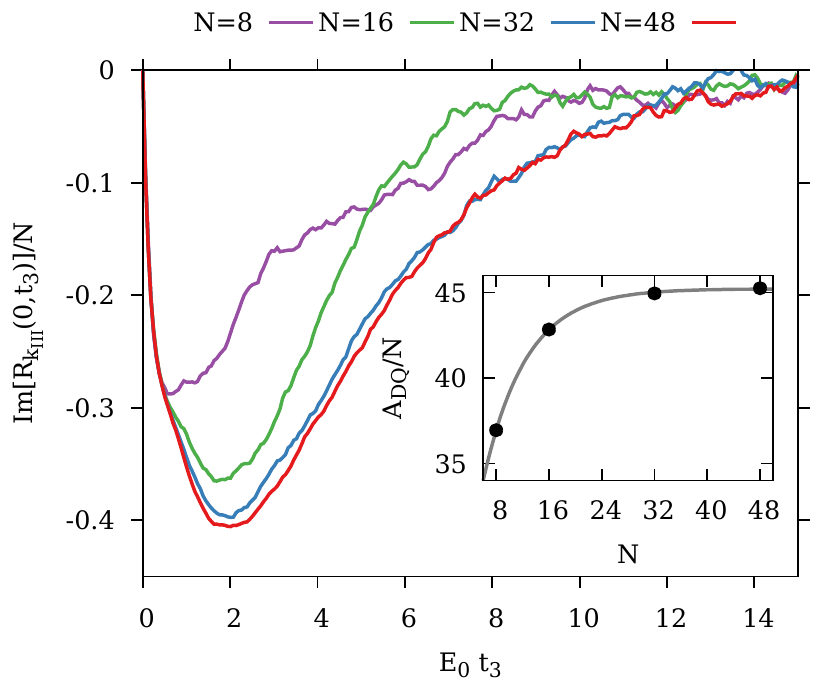}
\caption{Dependence of the imaginary part of the double quantum spectrum per atom $\im[R_{\vect{k}_{\text{III}}}(0,t_3)]/N$  for  $v_{\text{th}}=0$ and $t_2=0$ and as a function of $t_3$ and the number of atoms $N$ in the simulation box. The insert shows the dependence of the total double quantum spectrum amplitude  per atom  $A_{\text{DQ}}/N$ as a function of the number of atoms in the simulation box. The gray line is added as a guide to the eye. }
\label{fig:k3natoms}
\end{figure}
 We also show in Fig.~\ref{fig:k3natoms} the effect of the number of atoms included in the simulation box on the nonlinear signal. Specifically, we only show the imaginary part of the response function in the static case  $v_{\text{th}}=0$ for $t_2=0$ and as a function of $t_3$ but similar results are obtain in the case $t_2>0$ and  $v_{\text{th}} > 0$ (data not shown). In Fig.~\ref{fig:k3natoms}, the insert shows the size dependence of the total DQS signal per atom $A_{\text{DQ}}/N$.  As one expects, the nonlinear signal per atom becomes constant for large values of $N$. In addition, our simulations show that the collective behavior plays an important role in the DQS signal and is not limited to two atoms. In particular, a  value of the number of atoms larger than $N\geqslant 32$ is necessary to obtain converged results.\par
Our simulations show that the nonlinear signal is directly proportional to the number of atoms in the system. As noted earlier the total spectrum amplitude $A_{\text{DQ}}$ is completely independent of the energy scale $E_0$. Therefore $A_{\text{DQ}}$ only depends on the density through the number of atoms and therefore should vary linearly with the density $n$. However, as it will be discussed in sec.~\ref{sec:doppler} this is no longer valid if we include other broadening mechanisms such as Doppler broadening.
\section{Two-body approximation}
\label{sec:twobody}
In order to discuss our numerical results, we will derive expressions for the linear and nonlinear signals using the two-body approximation which considers only the interaction between pairs of nearest neighbor atoms and offers a correct description of the response functions at short times. This approximation was introduced in the context of atomic vapors in Ref.~\citenum{Leegwater:1994aa} to describe the effect of self-broadening on the linear absorption spectrum and we offer here an alternative derivation which we will then use to describe the DQS signal.\par
At short times, the electronic wavefunction will only delocalize between nearest neighbor atoms which can be considered as fixed. Therefore, the linear response function can be approximated by the average response of a single pair of atoms. The linear response function is then written
\begin{equation}
R_{\alpha\alpha}(t) \approx \int   R_{\alpha\alpha}^{(2)}(t;\vect{r}) \ P(\vect{r}) \ud^3 \vect{r},
\end{equation}
where $R_{\alpha\alpha}^{(2)}(t;\vect{r})$ is the response of a given pair of nearest-neighbor atoms located at a distance $\vect{r}=\vect{r}_2-\vect{r}_1$. Using the basis defined by Eq.~(\ref{eq:basismol}), the pair response function is given by
\begin{equation} 
R^{(2)}_{\alpha\alpha}(t;\vect{r}) =  \bra{+,\alpha}U(t) \ket{ + , \alpha} 
 =    e^{-\ii \omega_0 t }  \bra{\alpha} e^{-\ii \hat{J}(\vect{r}) t} \ket{\alpha},
 \label{eq:linearR2}
\end{equation}
where $\hat{J}(\vect{r})$ is the $3\times3$ matrix  given by Eq.~(\ref{eq:Jab}) which can be easily diagonalized analytically. We  obtain
\begin{equation} 
  \bra{\alpha} e^{-\ii \hat{J}(\vect{r}) t} \ket{\alpha}
 = e^{-\ii J(r) t}\left( 1 - \hat{r}_{\alpha}^2 \right) + e^{ 2\ii J(r) t } \hat{r}_\alpha^2,
 \label{eq:twobodypropagator}
\end{equation}
where $J(r) = E_0 / r^3$.
Performing the orientational average, the linear response function can be expressed as
\begin{equation}
R_{\alpha\alpha}(t) = e^{-\ii \omega_0 t} \frac{E_0}{3} \int_0^\infty \frac{\ud J}{J^2} e^{-E_0/J} \left[ e^{2\ii J t} + 2 e^{-\ii J t}  \right].
\label{eq:fullLA2body}
\end{equation}
We introduce the function $S(t) = 1 - e^{\ii \omega_0 t} R(t)$. It is written
\begin{equation}
S(t) = \frac{E_0}{3} \int_0^\infty \frac{\ud J}{J^2} e^{-E_0/J} \left[ 3 - e^{2 \ii J t} - 2e^{-\ii J t}\right].
\end{equation}
This integral can be rewritten using the change of variable $x= J t$
\begin{equation}
S(t) = \frac{E_0 t}{3} \int_0^\infty \frac{\ud x}{x^2} e^{-E_0 t/x} \left[ 3 - e^{2\ii x} - 2e^{-\ii x} \right].
\end{equation}
In the short time limit, we can keep only the linear term in the function $S(t)$ and neglect the term $e^{-E_0 t/x}$ . The function $S(t)$ is written
\begin{equation}
S(t) \approx \frac{E_0 t}{3}\left( 2\pi + 2\ii \log 2\right).
\end{equation}
We can then approximate the linear response function as 
\begin{equation}
R_{\alpha\alpha}(t) \approx e^{-\ii \left( \omega_0 + \Delta_S\right) t - \Gamma_S t},
\label{eq:LA2body}
\end{equation}
where $\Delta_S$ and $\Gamma_S$ are given by 
\begin{align}
&\Delta_S =  \frac{2}{3}\log 2 E_0, \\
&\Gamma_S = \frac{2\pi}{3} E_0,
\end{align}
which is exactly the expressions obtained by Leegwater and Mukamel using the two-body approximation.\cite{Leegwater:1994aa} In Eq.~(\ref{eq:LA2body}), the two-body approximation seems to give a Lorentzian lineshape, however this approximation is only valid at short times in the linear regime for two reasons. First, we have neglected the three-body contributions and many-body contributions in general. Second, we also did not take into account the complete time dependence of the two-body contribution as in Eq.~(\ref{eq:fullLA2body}) and only kept the linear contribution. As a result, the complete linear response does not give a Lorentzian lineshape.
\par
We now turn to the use of the two-body approximation to describe the DQS signal. We assume that the total nonlinear signal results from a set of $N/2$ pairs of nearest-neighbor atoms.  
\begin{equation}
R^{\vect{k}_{\text{III}}}_{xxxx}(t_2,t_3) \approx \frac{N}{2}  \int   R_{xxxx}^{(2)}(t_2,t_3;\vect{r}) \ P(\vect{r}) \ud^3 \vect{r},
\end{equation}
where $R_{xxxx}^{(2)}(t_2,t_3;\vect{r})$ is the nonlinear response function from a pair of nearest neighbor atoms.  As we are only considering a pair of atoms the two-exciton states are composed of a single state $b^\dagger_1b^\dagger_2\ket{g}$ and its propagation is therefore trivial. The nonlinear response function for the pair of atoms is then written
%
\begin{equation}
R^{(2)}_{xxxx}(t_2,t_3;\vect{r}) =  2 e^{-\ii \omega_0 \left(2  t_2 + t_3 \right) }  \sum_{i,j=1,2}  \left(  \bra{i x} e^{-\ii V_{12}(\vect{r}) t_3 } \ket{j x} - \bra{i x} e^{ \ii V_{12}(\vect{r}) t_3 } \ket{j x} \right), 
\label{eq:R2atoms}
\end{equation}
%
where $\ket{i x} = b_{i x}^\dagger \ket{g}$ is the one-exciton basis. Eq.~(\ref{eq:propagator}) shows that the two-body propagator is real valued for $i=j$ while it is purely imaginary in the case $i\neq j$. Therefore, the terms $i=j$ in Eq.~(\ref{eq:R2atoms}) vanish due to the interference between the two Liouville pathways defining DQS. The expression of the nonlinear response function is then written
%
\begin{equation}
R^{(2)}_{xxxx}(t_2,t_3;\vect{r}) =  8  e^{-\ii \omega_0 \left(2  t_2 + t_3 \right) } \bra{1 x} e^{-\ii V_{12}(\vect{r}) t_3 } \ket{2 x}, 
\end{equation}
%
This is an important result as it gives us a direct interpretation of DQS and how it is sensitive to the interatomic interaction. This equation shows that the signal is directly proportional to the transition amplitude of an electronic excitation between two different atoms. It is in fact a direct measurement of the delocalization of the electronic excitation between atoms due to dipole-dipole interactions. Note that this interpretation is only valid within the two-body approximation and gives us a physical interpretation of this spectroscopic technique only at short times.\par
For two atoms, the expression of the matrix elements of the propagator defined by Eq.~(\ref{eq:propagator}) can be simplified and we obtain
\begin{equation}
\bra{1 \alpha} e^{-\ii V_{12} (\vect{r}) t } \ket{2 \alpha} =  \left(1 - r_{\alpha}^2  \right)  \ii \sin J t  -r_{\alpha}^2  \ii  \sin 2Jt.
\end{equation}
Performing the orientational average, we obtain the expression for the DQS signal
\begin{equation}
R^{\vect{k}_{\text{III}}}_{xxxx}(t_2,t_3) \approx -\ii  \frac{4}{3} N E_0 e^{-\ii\omega_0 (2 t_2  + t_3) } \int_0^{\infty} \frac{\ud J}{J^2} e^{-E_0/J} \left( 2 \sin J t_3 - \sin 2J t_3 \right)
\end{equation} 
%
Using the change of variable $x=Jt$ this integral can be written
%
\begin{equation}
R^{\vect{k}_{\text{III}}}_{xxxx}(t_2,t_3) \approx - \ii  \frac{4}{3} N  e^{-\ii\omega_0 (2 t_2 + t_3) } E_0 t_3 \int_0^\infty \frac{\ud x}{x^2}\left( 2\sin x - \sin 2 x\right)  e^{-E_0 t_3/x}.
\end{equation}
%
In the short time limit we can neglect the term $e^{-E_0 t_3/x}$ in the integral and we find
\begin{equation}
R^{\vect{k}_{\text{III}}}_{xxxx}(t_2,t_3) \approx - \ii  \frac{8 \log 2 }{3} N e^{-\ii\omega_0 (2 t_2 + t_3)} E_0 t_3.
\label{eq:shorttime1}
\end{equation}
This equation shows that the two-body approximation only gives the initial increase of the nonlinear signal as a function of time $t_3$. It does not allow us to describe the decay of the signal as a function of the times $t_2$ and $t_3$. This can be easily understood when considering the two-exciton manifold which in the two-body approximation is only composed of a single state. Therefore, no decoherence can be observed.  To properly take into account the decay of the DQS signal, one needs to include many-body contributions. In Fig.~\ref{fig:k3timet3}, we compare our numerical results with the short time approximation as given by Eq.~(\ref{eq:shorttime1}). It shows that the initial increase of the imaginary part of the nonlinear response function is perfectly reproduced by the approximation, confirming the validity of the two-body approximation to describe the dynamics at short times.
%
\section{Effect of Doppler broadening}
\label{sec:doppler}
Up to now, in describing the spectral lineshapes, we have only considered self-broadening and neglected the effect of Doppler broadening. However, at a density lower or close to $n_1$, Doppler broadening should be included as it will be the main line broadening mechanism. To describe this effect we need to take into account the spatial phase of the electromagnetic field which can be included explicitly in the definition of the lowering operators $D_x$  
\begin{equation}
D_x(t) = \sum_i b_{ix}(t) e^{-\ii \vect{k}_L\cdot \vect{r}_i(t) }.
\end{equation}
To separate from the response functions defined earlier, we will denote with a tilde the response functions that include the effect of Doppler broadening. For example, the expression of the first Liouville pathway including Doppler broadening can be written
%
\begin{equation}
\tilde R^{(3a)}_{xxxx}(t_2,t_3) =  \sum_{i,j,k,l}  e^{-\ii \vect{k}_L\cdot \left( \vect{r}_l + \vect{r}_k - \vect{r}_j(t_2) - \vect{r}_i(t_2+t_3)   \right) }    \bra{g} b_{i x} U(t_2+t_3,t_2) b_{jx}  U(t_2,0) b_{kx}^\dagger b_{l x}^\dagger  \ket{g}. 
\end{equation}
%
A similar equation applies for the second Liouville pathway. We will use here the two-body approximation and only consider two atoms in the expression of $\tilde{R}^{(3a)}_{xxxx}(t_2,t_3)$. We obtain 
%
\begin{equation}
\tilde{R}^{(3a)}_{xxxx}(t_2,t_3)  = 2 \sum_{i,k\neq j} e^{-\ii \omega_0 (2t_2+t_3) }  e^{-\ii \vect{k}_L\cdot \left( \vect{r}_j + \vect{r}_k - \vect{r}_k(t_2) - \vect{r}_i(t_2+t_3)   \right) } 
\bra{ix}  e^{-\ii V_{12} t_3}  \ket{j x}. 
\end{equation}
%
As described previously, when considering the contribution of both Liouville pathways, only the terms $i\neq j$ remain. The nonlinear response function is then written
%
\begin{equation}
\tilde{R}^{(2)}_{xxxx}(t_2,t_3;\vect{r}) =  \sum_{i,j=1,2} 8  e^{-\ii \omega_0 \left(2  t_2 + t_3 \right) }  
    e^{-\ii \vect{k}_L\cdot \left( \vect{r}_{ji}- \vect{v}_i(2 t_2 + t_3 )  \right) } \bra{i x} e^{-\ii V_{12}(\vect{r}) t_3 } \ket{j x}, 
\end{equation}
%
where $\vect{r}_{ji}= \vect{r}_j - \vect{r}_i$. Finally, using the fact that  $n\gg n_0$, we can neglect the term $\vect{k}_L\cdot \vect{r}_{ji} $, and by averaging over the Maxwell-Boltzmann distribution for the velocities we find
\begin{equation}
\tilde{R}^{\vect{k}_{\text{III}}}_{xxxx}(t_2,t_3) = R^{\vect{k}_{\text{III}}}_{xxxx}(t_2,t_3) \times e^{-\frac{1}{2} k_0^2 v_{\text{th}}^2 \left( 2 t_2 + t_3\right)^2}.
\label{eq:dopplerbroadening}
\end{equation}
This equation shows that within the two-body approximation, the Doppler effect induces an inhomogeneous line broadening of the DQS signal that can be included using a simple Doppler shift. Since the phase of the double quantum spectrum is given by $\omega_0 (2 t_2 + t_3 )$ the Gaussian fluctuations of this phase factor due to the Doppler frequency shift give directly the Gaussian broadening function in Eq.~(\ref{eq:dopplerbroadening}). However, this interpretation is only valid because within the two-body approximation, the Doppler contribution arises from the velocity shift of a single atom. This can be understood by the following. Initially, the interaction with the two first pulses creates a doubly excited state over two neighboring atoms. The third pulse destroys one of the excitation on one of the atoms and the remaining excitation gets transferred onto the same atom again which is then destroyed by the last pulse. 
Without the two-body approximation, e.g. by including many-body contributions, a more complex spectral lineshape due to the Doppler effect should arise.\par
Combining Eqs.~(\ref{eq:shorttime1}) and~(\ref{eq:dopplerbroadening}) gives the DQS signal in the time domain when Doppler broadening is dominant compared to self-broadening. The signal in the frequency domain is then written
%
\begin{equation}
S_{\text{DQ}}(\omega_2,\omega_3) = \frac{8 \log 2}{3} N E_0 \tau^3 G\left( \left(\omega_2/2-\omega_0\right)\tau, \left(\omega_3-\omega_0\right)\tau \right), 
\end{equation}
%
where $\tau=1/(k_0 v_{\text{th}})$ and where the function $G(x,y)$ is the integral
\begin{equation}
G(x,y) = \int_0^\infty \!\!\! \int_0^\infty  \!\!\! \ud u  \ud v\  e^{\ii u x + \ii v y } \ii v e^{-\frac{1}{2}\left(u+ v\right)^2 }.
\end{equation}
The integral $G(x,y)$ can be computed analytically but gives a cumbersome expression as a function of the Dawson integral.\cite{Abramowitz:1972qf} It is in fact much more convenient to evaluate it numerically by performing a fast Fourier transform.\par
\begin{figure}
\centering
\includegraphics{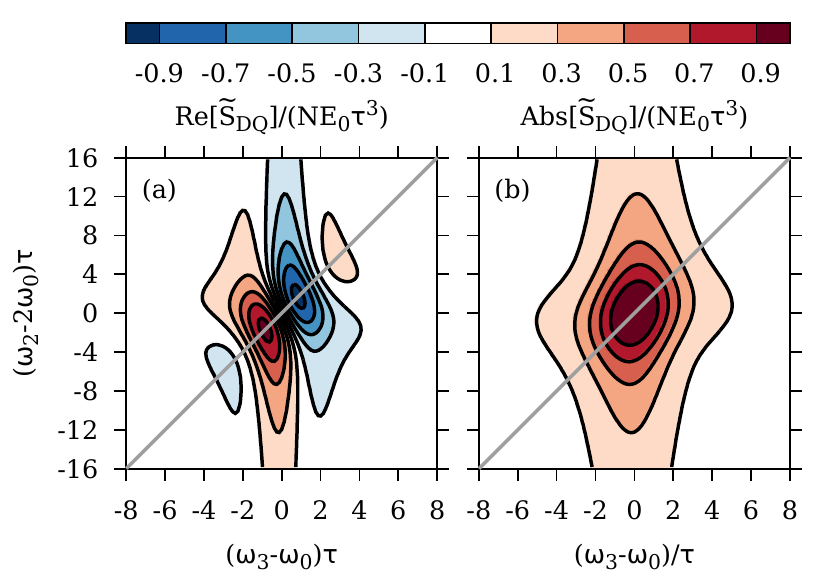}
\caption{Real part (a) and absolute value (b) of the DQS signal $S_{\text{DQ}} \left(\omega_2, \omega_3 \right)$ for an atomic vapor in the Doppler broadening regime.}
\label{fig:lineshape}
\end{figure}
In Fig.~\ref{fig:lineshape}, we show the real part and absolute value of the DQS signal $S_{\text{DQ}}(\omega_2,\omega_3)$ in the frequency domain as a function of the reduced shifts $\left(\omega_2-2\omega_0\right)\tau$ and $\left(\omega_3-\omega\right)\tau$ and in the Doppler broadening regime. The real part of the spectrum shows a clear dispersive behavior similarly as for the self-broadened spectrum. This is again a consequence of the linear increase of the response function in the time domain along the time $t_3$. Interestingly, the absolute value of the lineshape function shows a tilted character along the diagonal line $\omega_2 = 2\omega_3$ which was not observed earlier when we considered only self-broadening.  Therefore, the tilted shape of the DQS peaks can be directly attributed to inhomogeneous broadening as it has been pointed out before.\cite{Tollerud:2016aa,Lomsadze:2018aa,Lomsadze:2020aa}\par
In order to simulate the DQS over a large span of densities and interpolate between Doppler broadening and self-broadening, we will assume that the result given by Eq.~(\ref{eq:dopplerbroadening}) captures the essential contribution of Doppler broadening and we will use this equation even outside the range of validity of the two-body approximation. We have computed the DQS signal for a potassium vapor as a function of the density $n$, using the parameters of Table~\ref{tab:potassium}. For a given density we have computed the two-dimensional DQS signal with the corresponding value of $v_{\text{th}}$ by interpolating the numerical DQS spectrum in the time-domain obtained in Sec.~\ref{sec:results}. We then obtain the final DQS spectrum including Doppler broadening by using Eq.~(\ref{eq:dopplerbroadening}) which is then used to compute the total spectrum amplitude from Eq.~(\ref{eq:IDQ}). Here as the spectral linewidth strongly depends on the density we have used $\Delta\omega$ equal $10$ times the FWHM of the linear absorption spectrum.\par
\begin{figure}
\centering
\includegraphics{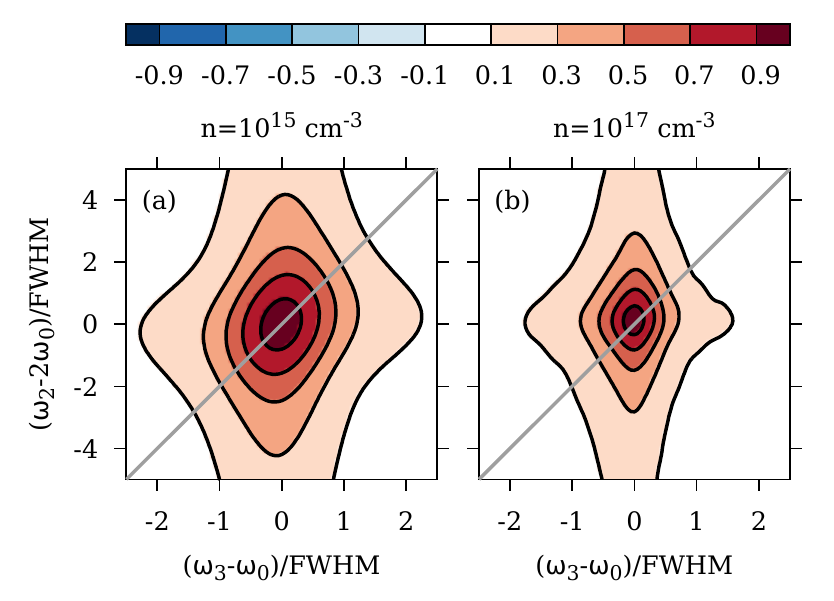}
\caption{Absolute value of the DQS signal  in the frequency domain $|S_{\text{DQ}}(\omega_2,\omega_3)|$  for a potassium vapor and for (a) $n=10^{15}$ cm$^{-3}$ and (b) $n=10^{17}$ cm$^{-3}$.}
\label{fig:potassium2D}
\end{figure}
As an example, we show in Fig.~\ref{fig:potassium2D} the absolute value of the two-dimensional DQS signal in the frequency domain in the case of a potassium vapor and for the densities $n=10^{15}$ cm$^{-3}$ and $n=10^{17}$ cm$^{-3}$. As the linewidth strongly depends on density, the DQS signal is plotted as a function of the reduced frequencies with respect to the FWHM of the linear absorption spectrum. For $n=10^{17}$ cm$^{-3}$ we recover a spectral lineshape similar to the one obtained in Fig.~\ref{fig:k3freq2d} in which Doppler broadening was not included and which shows no particular tilted character. For a density $n=10^{15}$ cm$^{-3}$ we recover a spectral lineshape corresponding to the Doppler regime obtained in Fig.~\ref{fig:lineshape} with a visible elongation of the peaks along the diagonal line defined by $\omega_2=2\omega_3$.\par
\begin{figure}
\centering
\includegraphics{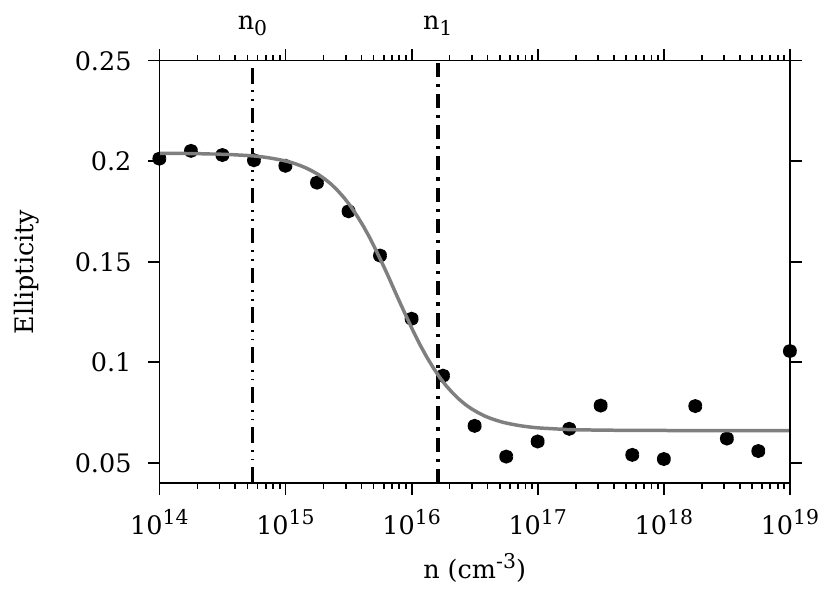}
\caption{Ellipticity of the DQS signal as a function of density for a potassium vapor. The gray line is added as a guide to the eye.}
\label{fig:ellipticity}
\end{figure}

 To analyze the spectral lineshape, and the tilted character of the DQS peak, we have measured the ellipticity $\epsilon$ as introduced in Ref.~\citenum{Lomsadze:2018aa}
\begin{equation}
\epsilon = \frac{a^2-b^2}{a^2+b^2},
\end{equation}
where $a$ and $b$ are the FWHM of the absolute value of the spectrum along the diagonal  line  ($\omega_2 = 2\omega_3)$ and along the antidiagonal line $\omega_2 = 4\omega_0 -2\omega_3$ respectively. The results are presented in Fig.~\ref{fig:ellipticity}. For very dense vapors, in the self-broadening regime the ellipticity has a value close to zero showing that the peaks are not tilted. As detailed in the Appendix, this is clearly a signature of the homogeneous character of self-broadening. As decreasing the density and entering the Doppler broadening regime, the peak acquires clearly a tilted behavior with an ellipticity close to 0.2. This is a signature of the inhomogeneous nature of Doppler broadening. Note that this value of the ellipticity is significantly smaller than the value measured by Lomsadze and Cundiff for a Rb vapor.\cite{Lomsadze:2018aa} However in our study we focused on dense atomic vapors with densities $n\gg n_0$ while in Ref.~\citenum{Lomsadze:2018aa} the densities are  below $n_0$. In this regime the nature of dipole-dipole interactions changes completely with a far-field contribution and retarded effects.\cite{James:1993aa} As approaching $n_0$ homogeneous broadening will switch from originating from the near-field contribution of dipole-dipole interactions resulting into self-broadening to its radiative, far-field contribution. As the homogeneous broadening will be modified, we should expect the overal lineshape to be modified as well.  In addition, our model does not take into account the fine and hyperfine structure of the atomic states. When including these interactions, the degeneracy of the atomic states are partially lifted and results in a modification of the vectorial nature of the dipole-dipole interactions and may have an impact on the spectral lineshape.\par
\begin{figure}
\centering
\includegraphics{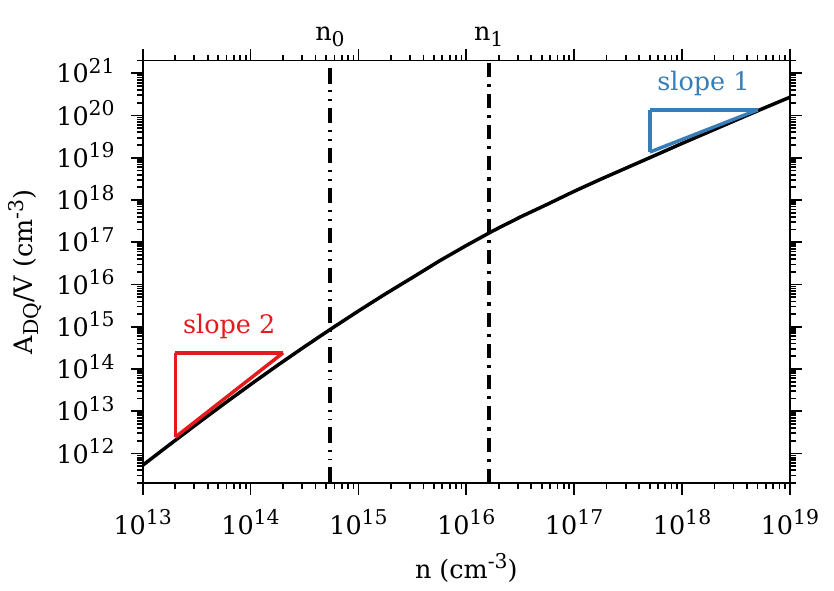}
\caption{Total DQS amplitude per unit volume as a function of density for a potassium vapor.}
\label{fig:doppler_k3}
\end{figure}

To show the effect of the density on the spectrum amplitude, we report in Fig.~\ref{fig:doppler_k3} the total spectrum amplitude $A_{\text{DQ}}$.  Our results show that for large densities, the nonlinear signal scales with the density as one expects and as we discussed at the end of Sec.~\ref{sec:results}. However, for lower densities and in particular when entering the Doppler broadening regime $n\lesssim n_1$ the nonlinear signal scales as the $n^2$.  This can be easily understood because in this regime the nonlinear signal scales linearly with the number of atoms and also with the strength of interaction $E_0$ which results in a $n^2$ scaling.\par
This effect is not limited to Doppler broadening but should also be visible when considering collisional broadening of an atomic vapor in a buffer gas. In this case, the homogeneous collisional broadening can be described by a simple Lorentzian lineshape and analytical results can be obtained. In the Appendix, we show that the DQS amplitude is then proportional to 
\begin{equation}
A_{\text{DQ}}  \propto N\frac{E_0}{\gamma},
\end{equation} 
where $\gamma$ is the main homogeneous broadening mechanism. At large densities, self-broadening dominates and $\gamma$ is proportional to $E_0$. Therefore, we can see that the signal is independent of $E_0$ and only depends on the number of atoms, such that 
\begin{equation}
\frac{A_{\text{DQ}}}{V}  \propto \frac{N}{V} = n.
\end{equation}
At low densities,  the collisional decay with the buffer gas dominates  and $\gamma$ is independent of the atomic density. In this case we have

\begin{equation}
\frac{A_{\text{DQ}}}{V}  \propto \frac{N}{V} \frac{E_0}{\gamma} \propto n^2,
\end{equation}
and we can see that the signal is proportional to the square of the density.
\section{Conclusions}
\label{sec:conclusions}
In this article, we have presented the theory of linear and nonlinear spectroscopy of dense atomic vapors specifically focusing on double quantum spectroscopy. We have investigated the effect of the thermal velocity on the linear and nonlinear signals and we have shown that the thermal velocity significantly contributes to the overall spectral lineshape. Using the two-body approximation, corresponding to the short time limit of the response functions, we show that double quantum spectroscopy measures directly the transition amplitude of the electronic excitation between atoms. Using the same approximation we are also able to explicitly include Doppler broadening. We show that in the case of pure Doppler broadening the DQS peaks are tilted along the diagonal $\omega_2 = 2 \omega_3$ while in the case of pure self-broadening the DQS peaks are not tilted. This confirms that the orientation of the peaks can be directly connected to homogeneous and inhomogeneous broadening. We also show that the total spectrum amplitude scales with the density when self-broadening dominates and scales as the square of the density when Doppler broadening dominates.\par
Our current study is limited to dense atomic vapors for densities $n\gg n_0 = k_0^3$. Performing measurements in this density regime is particularly difficult as it requires the use of a high-temperature vapor cell.\cite{Lorenz:2008uz} As a result, many recent DQS experiments have been performed in the very low density regime with $n\ll n_0$.\cite{Dai:2012qf,Gao:2016aa,Yu:2019aa,Liang:2021uj,Lomsadze:2018aa,Lomsadze:2020aa,Bruder:2015vk,Ames:2022tx} In this regime, when the excitation wavelength is much shorter than the average distance between atoms, the nature of dipole-dipole interaction changes completely.\cite{Ames:2022tx} In addition, the long wavelength approximation also does not apply anymore. We also have completely neglected the fine and hyperfine structure of the atomic state. As the atomic excitations have a fundamental vectorial nature, lifting degeneracies should have an important impact on the spectral lineshapes. Recently, DQS at very high resolution was performed using frequency comb technology showing the hyperfine structure of an atomic vapor.\cite{Lomsadze:2018aa,Lomsadze:2020aa} This underlines the necessity to include this effect explicitly as well. Consequently, our theoretical work cannot be directly compared to these experiments. Nevertheless, our study gives some important understanding of what DQS measures and how it is affected by the vapor density and Doppler broadening. In the future, we will extend this work to the low density regime and include the fine and hyperfine structure of the atomic states.
\begin{acknowledgments}
CF acknowledge the computational resources provided by the GRICAD infrastructure (https://gricad.univ-grenoble-alpes.fr), which is supported by Grenoble research communities. HL acknowledges the support by National Science Foundation (PHY-2216824).
\end{acknowledgments}

\appendix*
\section{Double quantum spectrum for Lorentzian lineshape}
\label{app:lorentz}
Using the two-body approximation, we have shown that the DQS signal is linear in the time $t_3$. However this approximation does not give the long time decay of the signal. It is possible to include self-broadening in an effective way by describing the long time decay of the signal using exponential functions. This can in principle describe other processes such as collisions if the atomic vapor is embedded in a buffer gas. We will therefore assume that the nonlinear signal in the time domain is written
%
\begin{equation}
R^{\vect{k}_{\text{III}}}_{xxxx}(t_2,t_3) = - \ii  \frac{8 \log 2 }{3} N e^{-\ii\omega_0 (2 t_2 + t_3 )} E_0 t_3 e^{-\gamma\left(2t_2+t_3\right)}, 
\end{equation}
%
where $\gamma$ is the decay rate. Using this function, the Fourier transform of the signal gives
\begin{equation}
S_{\text{DQ}}(\omega_2,\omega_3) = \frac{8 \log 2 N E_0/3}{\left(\omega_2 -2\omega_0+ 2\ii \gamma \right)\left(\omega_3 -\omega_0 + \ii \gamma\right)^2 },
\end{equation}
and its absolute value is then written
\begin{equation}
\left| S_{\text{DQ}}(\omega_2,\omega_3)  \right | =  \frac{8 \log 2 N E_0/3}{\sqrt{\left(\omega_2 -2\omega_0\right)^2 + 4 \gamma^2}\left(\left(\omega_3-\omega_0 \right)^2 + \gamma^2\right) }
\end{equation}
This spectrum lineshape is not tilted as the FWHM along the diagonal $a$ and antidiagonal $b$ are identical
\begin{equation}
a = b = 2\sqrt{2^{2/3}-1} \gamma\approx 1.53284 \gamma,
\end{equation}
which results in a vanishing ellipticity $\epsilon=0$.
The spectrum scales as $\omega_3^{-2}$ for large values of $\omega_3$ and it scales as $\omega_2^{-1}$ for large values of $\omega_2$. Therefore when defining the total spectrum amplitude as in Eq.~(\ref{eq:IDQ}), the integration over $\omega_2$  gives a logarithmic behavior as a function of $\Delta\omega$ which diverges for $\Delta\omega \rightarrow \infty$. For a Lorentzian lineshape the total spectrum amplitude can then be defined as
\begin{equation}
A_{\text{DQ}} = \int_{2\omega_0 - 2\Delta \omega}^{2\omega_0 + 2\Delta \omega} \int_{-\infty}^\infty \ud \omega_2  \ud \omega_3 \ | S_{\text{DQ}}(\omega_2,\omega_3) |,
\end{equation}
which gives
\begin{equation}
A_{\text{DQ}} =2 \pi  \frac{8 \log 2}{3} N \frac{E_0}{\gamma}\arcsinh\left(\frac{\Delta\omega}{\gamma} \right).
\end{equation}
As $\gamma$ defines the main broadening mechanism, $\Delta\omega$ should be proportional to $\gamma$. Therefore the total signal is thus proportional to 
\begin{equation}
A_{\text{DQ}}  \propto N\frac{E_0}{\gamma}.
\end{equation} 
\section*{Data Availability Statement}

The data that support the findings of this study are available from the corresponding author upon reasonable request.

\bibliography{biblio}

\end{document}